\documentclass[11pt]{article}

\pdfoutput=1

\usepackage[utf8]{inputenc}
\usepackage{fullpage}
\usepackage{graphicx}
\usepackage{hyperref}
\hypersetup{
    colorlinks=true,
    linkcolor=blue,
    filecolor=magenta,      
    urlcolor=cyan,
}

\title{A Note on Double Pooling Tests \\ (Preliminary version)}
\author{Andrei Z.~Broder \&\ Ravi Kumar \\
Google \\
Mountain View, CA\\[0.1cm]
{\tt broder@acm.org, ravi.k53@gmail.com}}

\date{\today}

\begin{document}

\maketitle

\begin{abstract}
We present double pooling, a simple, easy-to-implement variation on test pooling, that in certain ranges for the a priori probability of a positive test, is significantly more efficient than the standard single pooling approach (the Dorfman method). 
%The disadvantage of double pooling is that it is more sensitive to dilution-induced false negatives for two reasons: one physical: the pools used are larger; and one mathematical: a true positive sample will be missed if either of its two pools produces a false negative.
\end{abstract}

\section*{Introduction}

The concept of {\it test pooling} was apparently invented by Robert Dorfman~\cite{Dorfman} in 1943 who suggested that it would be more effective to test WW2 would-be recruits for syphilis by mixing the blood samples of several recruits and test the pool for antigens. If the pool tests negative then all the pool members are deemed healthy; otherwise, each member of the pool is tested separately.  A simple analysis (see next section) shows that for a given  probability $p$ that a recruit is infected there is an optimum pool size $s_1(p)$ that minimizes the expected number of needed tests.  The lower the $p$, the larger the $s_1(p)$ and the lower the expected number of tests required.  Dorfman's analysis has been further refined and generalized to deal with various problems such as false negatives~\cite{LLZA} and studied as part of the broad topic of Combinatorial Group Testing~\cite{CGT1, CGT2}.

Note that some recursive and adaptive approaches dear to computer scientists, such as binary search, often may not work for this problem: there are pragmatic limitations on (a) the size of the pool beyond which dilution results in too many false negatives; (b) the number of samples available from a given specimen; and (c) the total time required to produce an answer.  

Nevertheless the emergence of the COVID-19 pandemic and the cost and scarcity of tests for the underlying virus has revived an enormous interest in test pooling. For COVID-19 test pooling has been shown to be doable with pools as large as 64~\cite{Yelin} and is already in use in several countries including Germany~\cite{Frankfurt} and Israel~\cite{Yelin}.

\section*{Double pooling}

The purpose of this note is to propose a simple variation on Dorfman's approach that we call {\it double pooling}; for clarity, we refer to Dorfman's method as \emph{single pooling}.  Double pooling works as follows:  given a probability $p$ of a positive test, pick an optimal size $s_2(p)$ for the pool size.  (The optimal $s_2(p)$ is larger than the corresponding optimal $s_1(p)$ for single pooling.)  Divide the population to be tested into non-overlapping pools of size $s_2$ (the division is assumed to be random) \textit{twice}.  Thus, now every patient belongs to \textit{two} pools and is tested in two parallel rounds, $A$ and $B$.   For every patient if both the pools test positive then test the patient individually. Otherwise consider that patient cleared.  If the pool tests do not ever produce false negatives the algorithm is clearly correct.  (The false positives only reduce efficiency.)

It turns out the double pooling is particularly advantageous for $p < 2\%$ corresponding to testing a large population of asymptomatic patients but it is more efficient than single pooling even for $p=10\%$.

We will discuss double pooling in the next section in more detail, but to see its advantages and build an intuitive understanding we start with an example.  

Assume that $p=0.01112$.  It turns out $s_1=10$ is the optimal size for single pooling with this $p$ and results in an expected cost of $\approx 0.206$ tests/patient, a nice improvement over testing everyone.  However using double pooling the optimum $s_2(p)$ is 23 and the expected cost further declines to just $0.145$, an almost $30$\% improvement.  (These quantities will be obvious from our analysis later.)

At first blush these gains might seem surprising, but here is a quick-and-dirty computation:  Assume that we are testing 1000 patients and remember $p\approx0.011$ so we will posit we have exactly 11 positive patients. 
\begin{itemize}
  \item For single pooling, since $s_1 = 10$, we start with $100$ tests.  Assuming all positive cases end in separate pools (an upper bound) we will need to do another 110 tests to deal with all the suspicious cases, hence $210$ total tests (which is close enough to $206$.)
  
  \item For double pooling, since $s_2 = 23$, we  do twice 44 tests with 23 patients each (88 tests).  In each round  at most 11 tests  will come back positive raising suspicions about a total of  $22 \times 11 = 242$ healthy patients.  Thus a given healthy patient has probability $0.24$ to be a suspect in Round A and the same $0.24$ probability in round B. These are quasi-independent, hence the probability of being suspected twice is only $0.24 \times 0.24 = 0.0576$.  Thus we expect to have less than 
 $57$ ($\approx 0.0576 \times (1000-11)$) healthy patients that were in a positive pool in both rounds and will have to be retested.  In addition the 11 truly positive patients will be retested as well.    Thus the total number of tests is $88 + 57 + 11 = 156$ (which is reasonably close to the claimed $145$ since we overestimated and also because in fact we now cover $44 \times 23=1012$ patients).  
\end{itemize}

In conclusion, the ``magic"  of double pooling comes from a paradigm that has been observed in many other situations, e.g.,~Bloom filters~\cite{Bloom, DBLP:journals/im/BroderM03} and balanced allocations~\cite{DBLP:journals/siamcomp/AzarBKU99, DBLP:journals/tpds/Mitzenmacher01}.  Although the probability of being ``unlucky" in a given trial might be high, the probability of being unlucky in two or more independent trials decreases dramatically. 

\section*{Analysis}

% let $p \in (0, 1)$ denote the infection probability of a user, let $s$ denote the size of the pool, and let $k$ denote the number of pools a user participates.  Let $n$ be the total number of users.  

Consider the expected cost attributable to  one patient in the single pooling situation, where the size of the pool is $s$:
\begin{itemize}
    \item if the patient is positive, then the cost is $1/s + 1$ (the patient's share of the pool + their individual test);
    \item if the patient is negative, then the cost is $1/s + 1 \times \big(1-(1-p)^{(s-1)}\big)$  (the patient's share of the pool + their individual test \textit{iff} not all the other patients are healthy).
\end{itemize}

\noindent 
Since the probability of being positive is $p$, the total expected cost per patient in this case is 
\begin{equation} \label{tcsp}
p \left(1+\frac{1}{s}\right)+(1-p) \left(1-(1-p)^{s-1}+\frac{1}{s}\right)
= 
\frac{1}{s} + 1- (1-p)^s.
\end{equation}

To determine the pool size that minimizes the total for a given $p$, we take the derivative of the cost with respect to $s$:
\begin{equation} 
\frac{\partial}{\partial s} \left( \frac{1}{s} + 1- (1-p)^s \right)
=- (1-p)^s \ln(1-p)- \frac{1}{s^2},
\end{equation}
 and set it to $0$.
 
The solution of interest can be expressed in terms of the Lambert $W$ function%
\footnote{\url{https://en.wikipedia.org/wiki/Lambert_W_function}} 
namely 
\begin{equation}
s_1 
%= \left.
%2 W \left(\frac{1}{2} \sqrt{-\frac{1}{\log (1-p)}} \log (1-p)\right) \middle/ {\log (1-p)} \right.
= \left. 2 W\left( -\frac{1}{2} \sqrt{-\log(1-p)} \right) \middle/ \log(1-p)
\right. .
\end{equation}

\begin{figure}[h]
\centerline{
\includegraphics{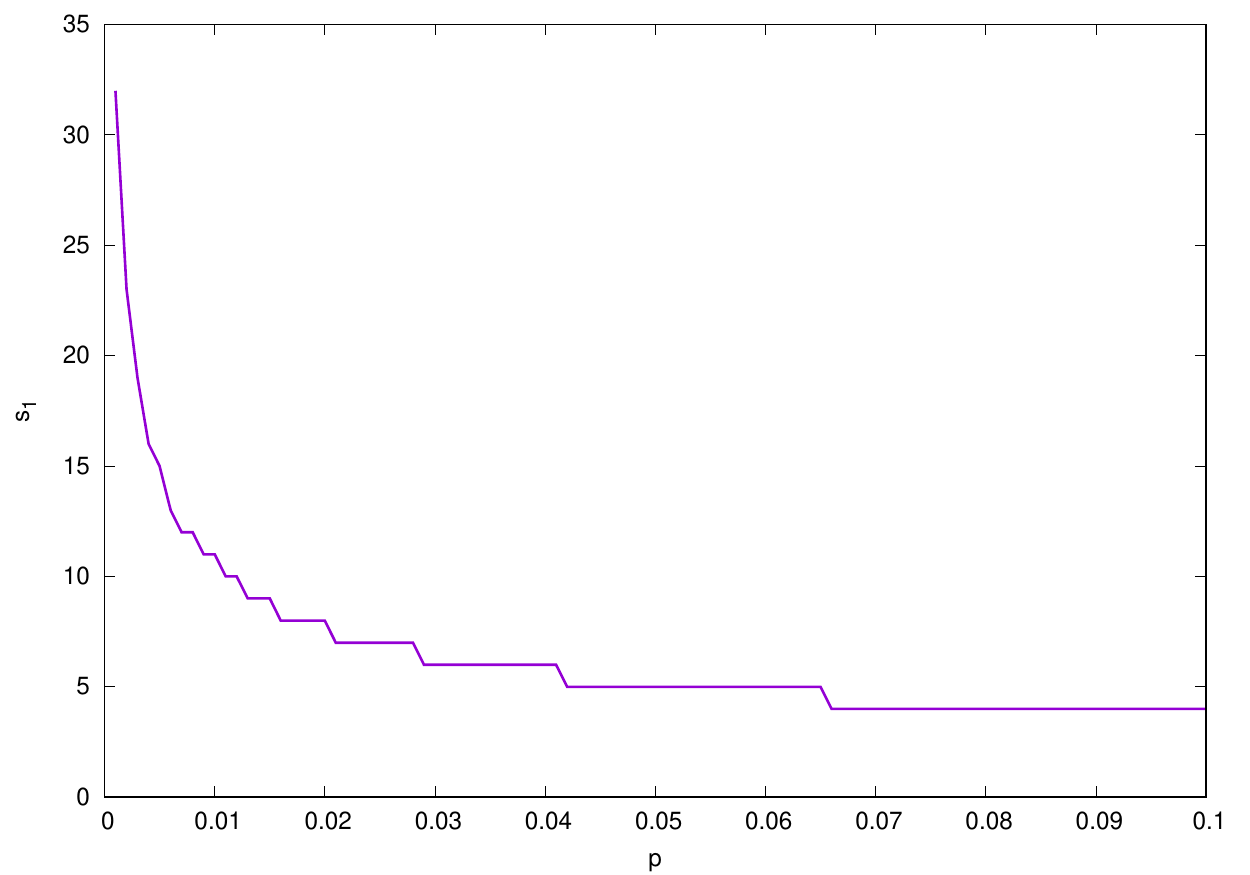}
}
\caption{Optimum $s_1$ as a function of $p$.  \label{fig:s1}}
\end{figure}

Let us define $p_{10}$ as the value of $p$ for which  the optimum $s_1$ is exactly 10.  It turns out that $p_{10}\approx0.01112$ which is the value we used in the introductory example.  More generally, Figure~\ref{fig:s1} shows the optimum integer $s_1$ as a function of $p$. 

\bigskip

Let us  turn to double pooling: now each patient will be assigned to two random pools each of size $s$.  A patient will be tested individually \textit{iff} both their pools test positive.  Again let us look at the expected cost induced by the testing of one patient:
\begin{itemize}
    \item if the patient is positive, then the cost is $2/s + 1$ (the patient's share of the two pools + their individual test);
    \item if the patient is negative, then the cost is $2/s + 1 \times \bigr(1-(1-p)^{(s-1)}\bigl)^2$  (the patient's share of the pools + their individual test \textit{iff} \emph{both} pools test positive).
\end{itemize}
Hence the total expected cost is
\begin{equation} \label{tcdp}
p \left(1+\frac{2}{s}\right)+q  \left(\bigl(1-q^{s-1}\bigr)^2+\frac{2}{s}\right)
=
\frac{2}{s} + p + q \bigl(1-q^{s-1}\bigr)^2,
\end{equation}
where for brevity $q$ stands for $1-p$.

As before, to determine the pool size that minimizes the total cost for a given $p$, we take the partial derivative of the cost with respect to~$s$:
\begin{equation} 
\frac{\partial}{\partial s} 
\left(
\frac{2}{s} + p + q \bigl(1-q^{s-1}\bigr)^2
\right)
=  - \frac{2}{s^2} - 
2 q^{s} \bigl(1-q^{s-1}\bigr) \ln q,
\end{equation}
and set it to $0$.

\begin{figure}[h]
\centerline{
\includegraphics{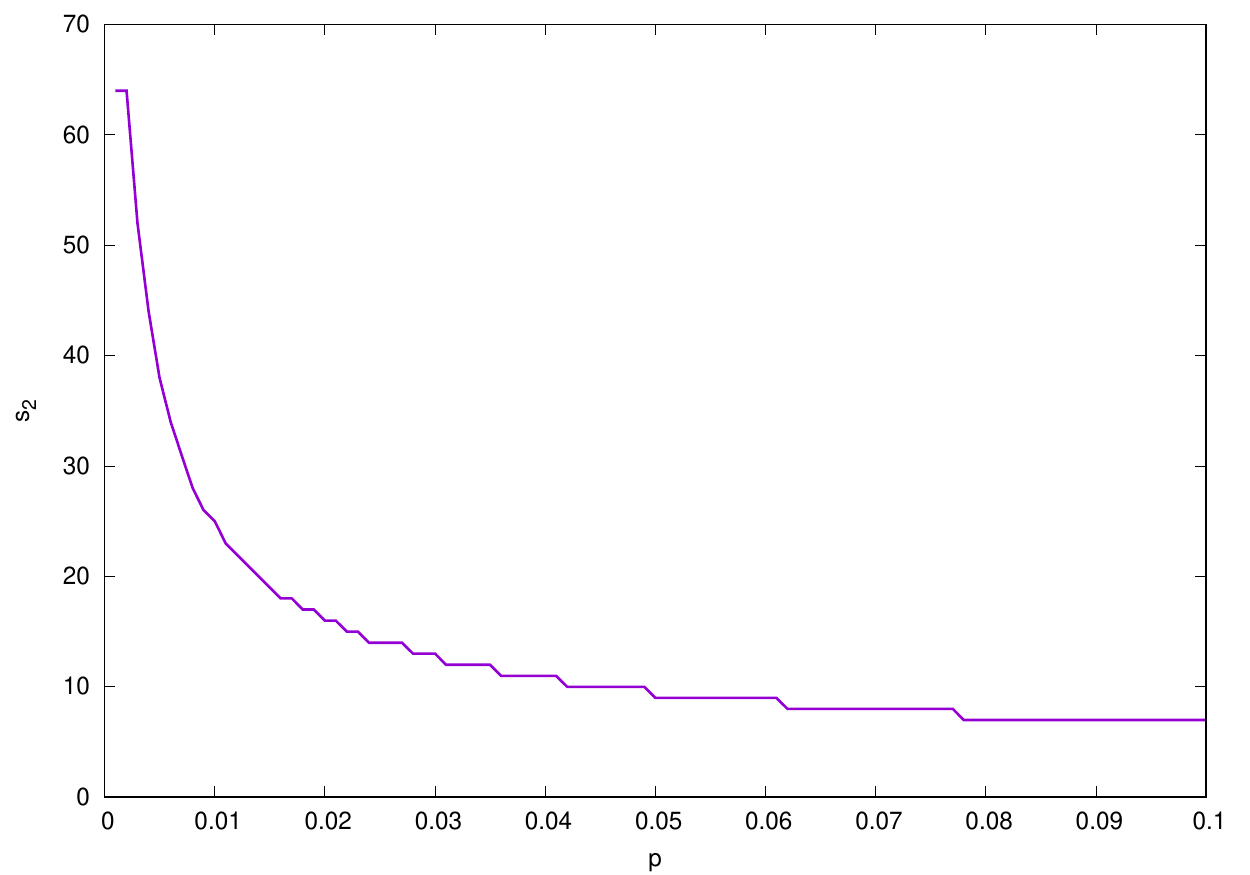}
}
\caption{Optimum $s_2$ as a function of $p$.  \label{fig:s2}}
\end{figure}

\begin{figure}[h]
\centerline{
\includegraphics{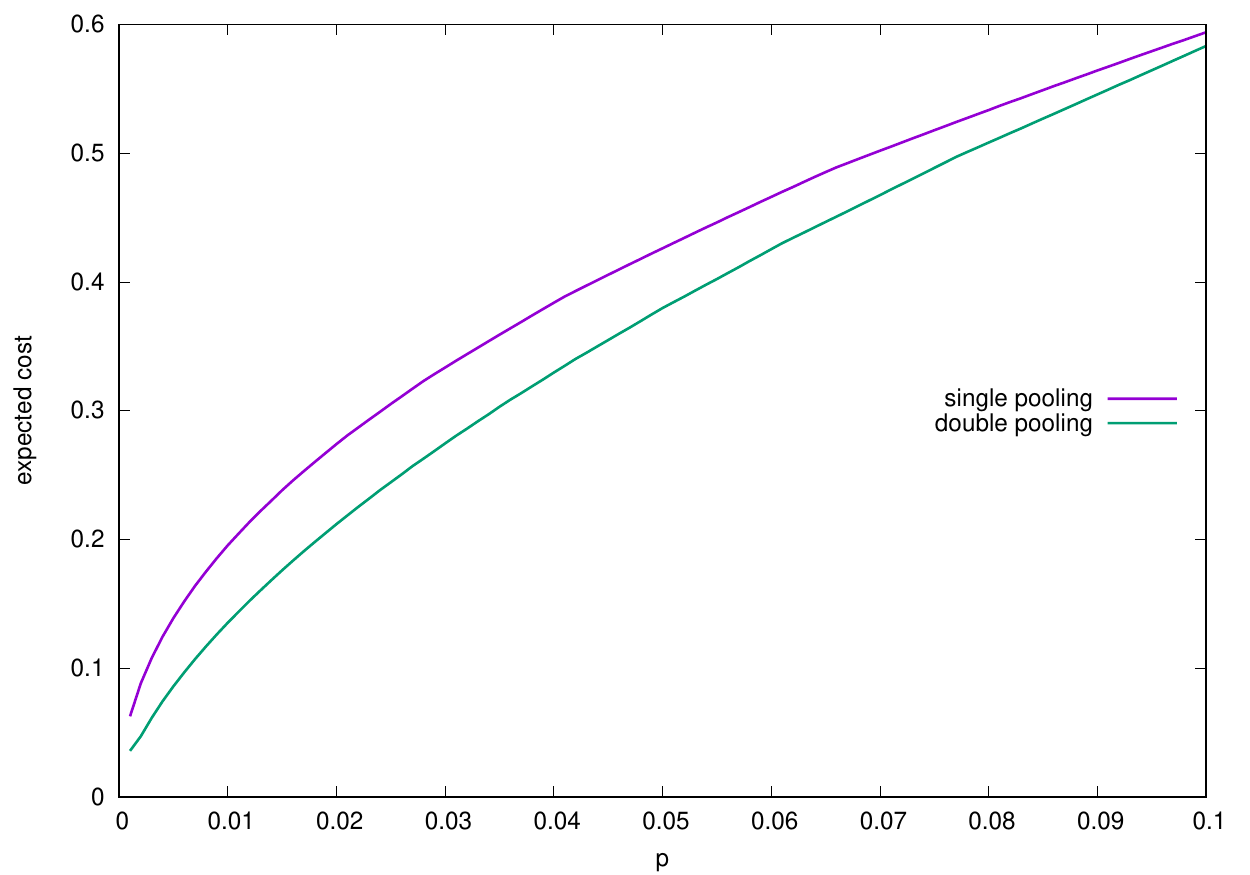}
}
\caption{Expected cost using the optimum $s_1$ and $s_2$ as a function of $p$.  \label{fig:c12}}
\end{figure}

This has to be solved numerically at each $p$.  Solving this at $p = p_{10} \approx 0.01112$ we see that the optimum $s_2 = 23$.  More generally, Figure~\ref{fig:s2} shows the optimum integer $s_2$ as a function of $p$ and Figure~\ref{fig:c12} shows the expected cost per patient tested as a function of $p$ for both single and double pooling using optimal integer values of $s_1(p)$ and $s_2(p)$.

\bigskip

\begin{figure}[h]
\centerline{
\includegraphics{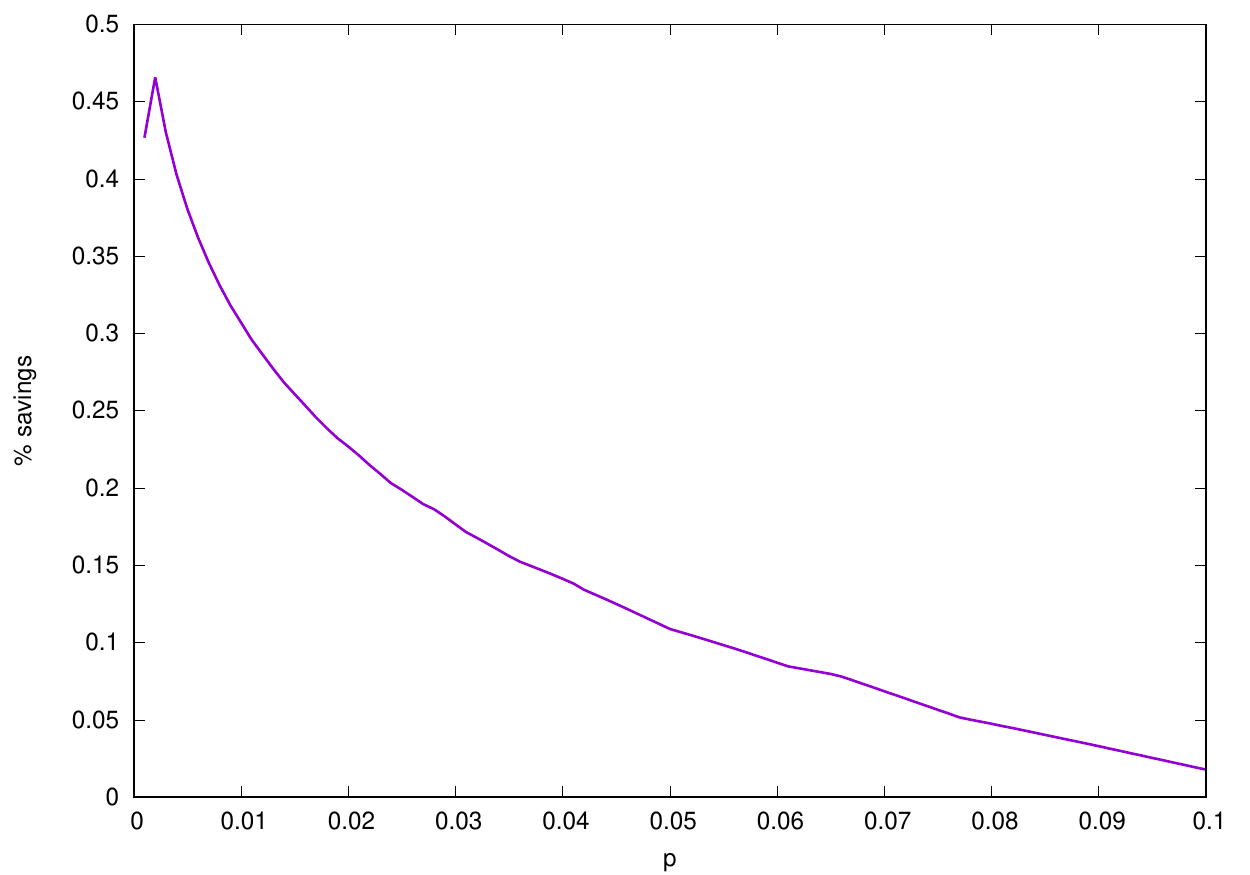}
}
\caption{\% savings of double pooling over single pooling as a function of $p$.  \label{fig:savings}}
\end{figure}

\bigskip

In principle we can generalize double pooling to  $k$-pooling, whereby each patient participates in $k$ independent pools in $k$ parallel rounds. The expected cost becomes
\begin{equation} \label{tckp}
\frac{k}{s} + p + q \bigl(1-q^{s-1}\bigr)^k.
\end{equation}
Depending on $p$ this can yield further improvements but they are probably impractical especially if $p$ gets larger. (With triple testing for $p=p_{10}$ and about 1000 samples  we would need only 128 tests with pools of size  36, and with quadruple testing only 122 tests with pools of size 47.).  Even more asymptotically efficient tests can be constructed~\cite{MezardToninelli, PoratRothschild}, but it is unclear if they can be practical.

\section*{Conclusions}
We presented \emph{double pooling}, a simple, easy-to-implement variation on test pooling, that in certain ranges for $p$, the  a priori probability of a positive tests, is significantly more efficient than the standard single pooling approach.  Figure~\ref{fig:savings} shows the percentage of savings of double pooling over single pooling as a function of $p$.  We can see that double pooling is particularly advantageous for $p$ below $2\%$  corresponding to large scale testing of asymptomatic patients, but is still at least $10\%$ better than single pooling all the way up to $p=5.4\%$.

Our analysis assumes sampling from an infinite distribution, but in practice, double pooling can be implemented  after accumulating a fairly small collection of samples.  (There is a small efficiency penalty due  to the correlation between rounds that we will discuss in the final version.)  The main disadvantage of double pooling is that it is more sensitive to dilution-induced false negatives for two reasons: one physical:  the pools used are larger; and  one mathematical: a true positive sample will be missed if either of its two pools produces a false negative.  We will discuss this further in the final version. 

We are reaching out to our colleagues in the medical field to find out whether double pooling is practically usable for COVID testing and will update our note with their feedback. 

Presently, there is an extraordinary flurry of activity  and independent work on group testing for COVID.  This includes an analysis of a single pooling method~\cite{Gollier} and a proposal based on binary search~\cite{Gossner}. It might well be the case that independent researchers have already obtained the same results presented here.  We are encouraging members of the community to send us their comments and feedback.

\subsection*{Acknowledgments}

We thank our colleagues Fernando Pereira and Tam\'{a}s Sarl\'{o}s for many  useful comments.

\bibliographystyle{alpha}
\bibliography{pooling}

\end{document}